# Realization of vortex pair and its application in optical tweezers

Zhongsheng Man,[1,2,*] Shuoshuo Zhang,[1] and Shenggui Fu[1]

[1] *School of Physics and Optoelectronic Engineering, Shandong University of Technology, Zibo 255000, China*
[2] *Optics Research Group, Delft University of Technology, Lorentzweg 1, 2628CJ Delft, The Netherlands*

As one fundamental property of light, the orbital angular momentum (OAM) of photon has elicited widespread interest. Here, we theoretically demonstrate that the OAM conversion of light without spin state can occur in homogeneous and isotropic medium when a line-variant locally linearly polarized (LVLLP) beam is strongly focused by a high numerical aperture (NA) objective lens. The high-NA objective lens here acts as a modulator that enables the spin-to-orbital OAM conversion of the two components of left and right circular vibrations of the input SDLLP beam spatially separated. Hence, partial conversion from linear state to conjugate OAM states takes place, resulting in helical phases with opposite directions for the longitudinal component of the two foci. Furthermore, such customized light field can be used to identify and separate chiral objects in optical tweezers.

It is well-known that a light beam can possess angular momentum (AM), in addition to linear momentum [1–11]. There are two categories of AMs including spin angular momentum (SAM) and orbital angular momentum (OAM). SAM is intrinsic and related to the vectorial nature of light, and it has two possible quantized values of $\pm\hbar$ depending on the handedness of circular polarization: $+\hbar$ per photon for a left-handed circularly polarized beam and $-\hbar$ per photon for a right-handed circularly polarized beam, where $\hbar$ is Planck's constant $h$ divided by $2\pi$ [1,2]. By contrary, OAM has both intrinsic and extrinsic terms, and the latter of which is coordinate dependent [5]. The intrinsic OAM, hereafter simply referred to as OAM, is related to the azimuthal dependence of optical phase. When a light beam possesses a vortex phase of $\exp(il\phi)$, it can carry an optical OAM of $l\hbar$ per photon, where $l$ is topological charge, indicating the repeating rate of $2\pi$ phase shifts azimuthally along the beam cross-section [3–11]. Such a vortex beam exhibits a helical wave-front and possesses a phase singularity at the beam center, resulting in a doughnut-shaped intensity profile [12,13].

Since the discovery of light's OAM [3], optical vortices have provided insights into the fundamental properties of light and lead to abundant applications, including micromanipulation [14,15], optical communication [16–21], super-resolution imaging [22–24], quantum information processing [25,26], and others. Great successes have been achieved in the creation and manipulation of optical OAM. Conventionally, OAM beams may be generated in various ways like spiral phase plate [27–30], computer-generated holograms [31,32], sub-wavelength gratings [33]. These techniques rely on introducing a phase discontinuity in the wave-front to generate beams with desired OAM modes. Generally, it is believed that polarization and phase are two relatively independent degrees of freedom (DoFs) of light that show little interaction. Nevertheless, under specific conditions, the intrinsic optical DoF of polarization also enables the manipulation of optical OAM states via the procedure refer to as spin-orbital-conversion (SOC) [34–43]. Such a kind of SOC provides a direction connection between SAM [circularly polarized state] and OAM and allows for a broadband manipulation of OAS states, in contrast to the aforementioned modulations that are generally wavelength dependent. For the traditional SOC, the mapping from SAM is limited to circular polarization (CP). The most general state of polarization (SoP) is elliptical polarization (EP). CP is only one extreme case of an infinite set of EPs, and the other is linear polarization (LP). Most recently, the conversion of arbitrary SAM states [the general elliptically polarized states] into states with independent values of OAM is achieved [44]. Such process requires the interaction of light with matter that is both optically inhomogeneous and anisotropic. As a counterpart, the achievement of such conversion is still a challenge in homogeneous and isotropic medium when the input optical field has no spin state [every light vibration is linearly polarized for each photon in the beam cross-section].

Here, we theoretically demonstrate that the OAM conversion of light without any spin state can indeed occur in homogeneous and isotropic medium when a line-variant locally linearly polarized (LVLLP) beam is strongly focused by a high numerical aperture (NA) objective lens. Based the vector diffraction theory, the analytical mode is built to calculate the three-dimensional electromagnetic field and Poynting vectors in the focal region of the proposed LVLLP beam. Through a high NA objective lens, the LVLLP beams can generate similar two foci with tunable distance between them controlled by input SoPs for all the transverse, longitudinal, and total fields. One is contributed by the input right-handed (RH) component while the other is contributed by the input left-handed (LH) component. The high-NA objective lens here acts as a modulator that enables the SOC of the two components of RH and LH circular vibrations of the input LVLLP beam spatially separated. Hence, partial conversion from linear state to conjugate OAM states takes place, resulting in helical phases with opposite directions for the longitudinal component of the two foci. Furthermore, such customized light field can be used to identify and separate chiral objects in optical tweezers.

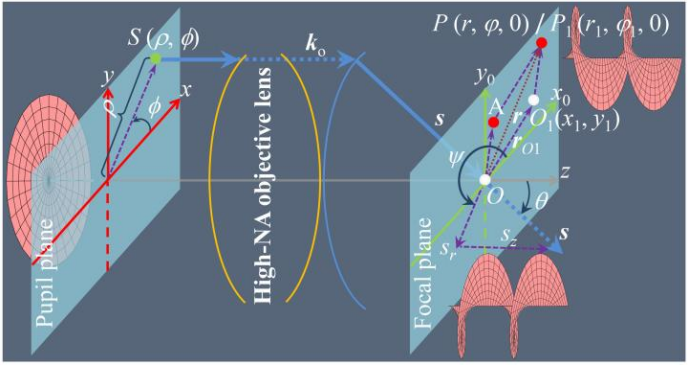

FIG. 1. Schematic of the concept for OAM conversion of light without spin state and geometry as well as coordinate system we followed in our calculations. The LVLLP plane beam, propagating along the $z$ axis is incident on a high-NA objective lens. The high-NA objective lens here acts as a modulator that enables the SOC of the two components of opposite phase modulated RH and LH circular vibrations of the input LVLLP beam spatially separated. Hence, partial conversion from linear state to conjugate OAM states takes place, resulting in helical phases with opposite directions for the longitudinal component of the two foci.

Polarization, as an intrinsic optical DoF, is one of the most salient features of light. In additional to the simplest and most fundamental homogenous SoPs, a light beam admit spatially inhomogeneous SoPs, which is the so-called vector optical field (VOF) [46–56]. Theoretically, a light with arbitrary locally linear SoP can be described as follows [46–56].

$$\mathbf{E}_o = A(x, y) \times \left[ \cos\delta(x, y)\hat{\mathbf{e}}_x + \sin\delta(x, y)\hat{\mathbf{e}}_y \right], \quad (1)$$

where $A$ represents the complex amplitude, and $\delta$ indicates the distribution of polarization, which can have arbitrary mathematical forms in theory. Apparently, the local SoP in the beam cross-section of the light denoted by Eq. (1) is linearly polarized, because the orthogonal base vectors in terms of $x$ and $y$ components are always in phase. Using the relations $\hat{\mathbf{e}}_x = \frac{1}{\sqrt{2}}(\hat{\mathbf{e}}_r + \hat{\mathbf{e}}_l)$ and $\hat{\mathbf{e}}_y = \frac{j}{\sqrt{2}}(\hat{\mathbf{e}}_r - \hat{\mathbf{e}}_l)$, equation (1) can also be represented in terms of orthogonal circularly polarized base vectors as

$$\mathbf{E}_o = \frac{A(x, y)}{\sqrt{2}} \times \left\{ \exp[-j\delta(x, y)]\hat{\mathbf{e}}_l + \exp[j\delta(x, y)]\hat{\mathbf{e}}_r \right\}, \quad (2)$$

where $\hat{\mathbf{e}}_l$ and $\hat{\mathbf{e}}_r$ are, respectively, the LH and RH circularly polarized unit vectors. As depicted in Eq. (2), any local linear vibration can be viewed as the superposition of two circular vibrations that have opposite handedness and carry opposite phase. So, linear polarization does not externally exhibit SAM arising from that it internally carries opposite SAMs, which implies that the OAM conversion of light without spin state is possible. For making such conversion dominant, the key is enabling the SOC of the two components of circular vibrations spatially separated. There is only one variable of $\delta$ can be used shown in Eqs. (1) and (2), hence it should be specially designed.

For achieving the OAM conversion from linear state in homogeneous and isotropic medium, a high-NA objective lens is introduced, as depicted in Fig. 1. Tight focusing is highly demanded ranging from microimaging, optical micromanipulation, high-density storage, and others. In such a system, the optical field in the image space should be analyzed using vectorial diffraction theory [57], due to the contribution of input polarization cannot be neglected, in contrast to the Huygens Fresnel's principle. According to Richards and Wolf vectorial diffraction theory, the electric field near focus of arbitrary polarized beam is given by a diffraction integral over the vector field [57,58]:

$$\mathbf{E}_i = \frac{-ik_1}{2\pi} \iint_\Omega \mathbf{a} \times \exp[ik_1(\mathbf{s}\cdot\mathbf{r})]d\Omega, \quad (3)$$

where $k_1$ is the vacuum wave number in image space, $\mathbf{a}$ is the strength factor that is related to the object space electric field, $\mathbf{s}= (\sin\theta\cos\psi, \sin\theta\sin\psi, \cos\theta)$ is the unit vector along a typical ray in image space, $\mathbf{r}$ is the radius vector of arbitrary point $P$ ($r$, $\varphi$, $z$) in image space, $\Omega$ is the solid angle formed by all the geometrical rays which pass through the exit pupil of the system.

When the paraxial focus is located at the origin $O$ (0, 0, 0), for an arbitrary point $P$ near the paraxial focus in the image region and consider $\psi = \pi + \phi$, where $\phi$ is the azimuthal angle with respect to $x$ axis.

$$\mathbf{s}\cdot\mathbf{r} = -r\sin\theta\cos(\phi-\varphi) + z\cos\theta. \quad (4)$$

By contrast, when paraxial focus is shifting to another position $O_1$ ($x_1$, $y_1$, 0) in the focal plane, with radius vector $\mathbf{r}_{O1} = (x_1, y_1, 0)$, considering the spatial translation invariance of focus, $\mathbf{s}\cdot\mathbf{r}$ in Eq. (3) should be modified as

$$\begin{aligned}\mathbf{s}\cdot(\mathbf{r}-\mathbf{r}_{O1}) &= \mathbf{s}\cdot\mathbf{r} - \mathbf{s}\cdot\mathbf{r}_{O1} \\ &= -r\sin\theta\cos(\phi-\varphi) + z\cos\theta + \\ & \quad x_1\sin\theta\cos\phi + y_1\sin\theta\sin\phi \end{aligned} \quad (5)$$

Substituting Eq. (5) into Eq. (3), then

$$\mathbf{E}_i = \frac{-ik_1}{2\pi} \iint_\Omega \mathbf{a}_1 \times \exp[ik_1(\mathbf{s}\cdot\mathbf{r})]d\Omega, \quad (6)$$

where $\mathbf{a}_1 = \mathbf{a}\exp[-ik_1(\mathbf{s}\cdot\mathbf{r}_{O1})]$. Physically, Eq. (6) indicates that the input field in object space is modulated by an additional phase, compared with the situation that in Eq. (3). Based on Eqs. (5) and (6), the corresponding phase modulations in object space can be derived, when the high-NA objective lens obeying sine condition, as

$$\Delta = \frac{2\pi \cdot \mathrm{NA} \cdot x_1}{n_1 \rho_o \lambda} x + \frac{2\pi \cdot \mathrm{NA} \cdot y_1}{n_1 \rho_o \lambda} y = \frac{2\pi \cdot h}{\rho_o} x + \frac{2\pi \cdot v}{\rho_o} y, \quad (7)$$

where NA and $n_1$ is the numerical aperture of objective lens and the refractive index in image space that we take as 0.95 and 1 in the following calculations, respectively, $\rho_0$ and $\lambda$ denotes, respectively, the radius and wavelength of input optical field, $h$ and $v$ are two introduced parameters, which represent, respectively, horizontal and vertical indexes, that is similar to topological charge $l$ of optical vortex.

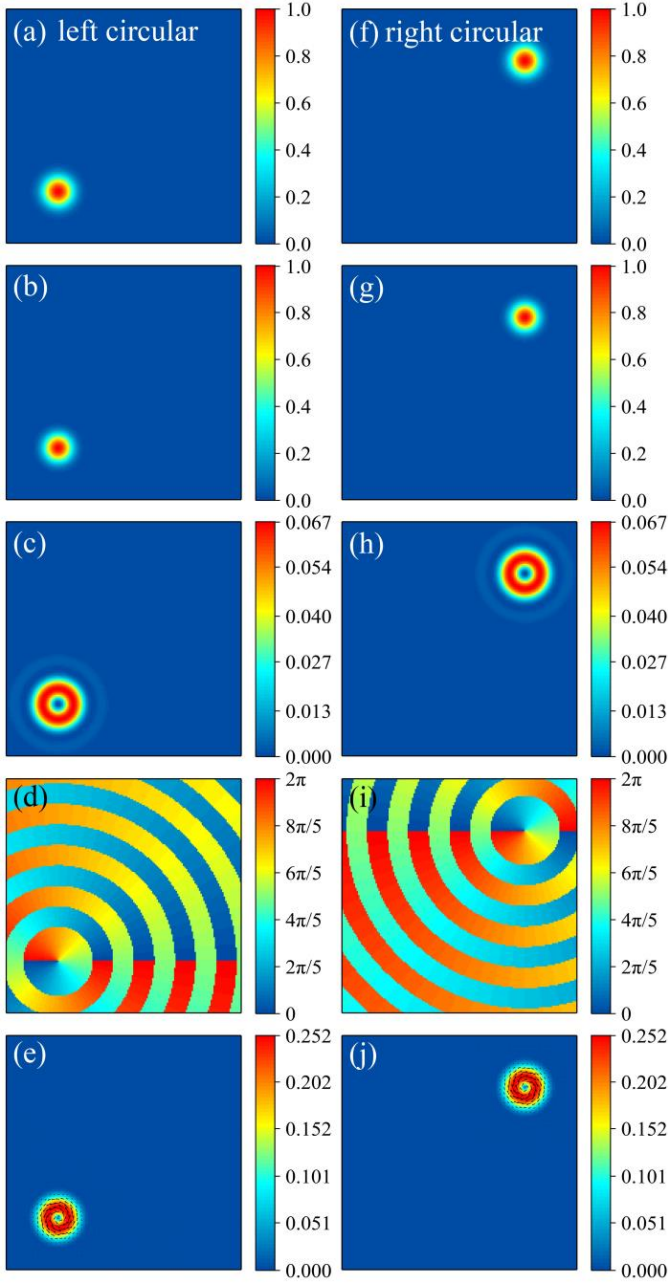

FIG. 2. Simulated electric field and energy flow of tightly focused opposite phase modulated LH (left column) and RH (right column) circularly polarized $LG_{(0,0)}$ beams in the focal plane. (a),(f) Total intensity distribution. (b),(g) Intensity distribution of the transverse component. (c),(h) Intensity distribution of the longitudinal component. (d),(i) Phase distribution of the electric field of the longitudinal component. (e),(j) Energy flow of the transverse component, the direction of which is shown by the black arrows. The size for each image is $6\lambda \times 6\lambda$.

Theoretically, when $\delta$ in Eq. (2) has the same form as $\Delta$ in Eq. (7), the components of opposite phase modulated RH and LH circularly polarized optical fields should be focused to different locations at $(x_1, y_1, 0)$ and $(-x_1, -y_1, 0)$, respectively, which are origin of symmetry. As a result, spatially separating the two components can be achieved by specially designing the parameter $\delta$. To verify such phenomenon, we employ the circularly polarized Laguerre-Gaussian laser mode $LG_{(l,p)}$ [4], where $l$ and $p$ are, respectively, the numbers of interwined helices known as the topological charge and the additional concentric rings. For considering the most fundamental case of $l = p = 0$ and $\delta = \Delta$, the strength factor $\mathbf{a}_1$ in Eq. (6) for the input two components in term of $\frac{A(x,y)}{\sqrt{2}}\exp[-j\delta(x,y)]\hat{\mathbf{e}}_l$ and $\frac{A(x,y)}{\sqrt{2}}\exp[j\delta(x,y)]\hat{\mathbf{e}}_r$ in Eq. (2) can be expressed, respectively, as

$$\mathbf{a}_1 = fl_1\sqrt{\cos\theta}\mathbf{M_e}, \quad (8)$$

$$l_1 = \frac{1}{\sqrt{2}}\exp\left[-\left(\beta\frac{\sin\theta}{\sin\alpha}\right)^2\right]e^{\left[\mp j\frac{2\pi\sin\theta}{\sin\alpha}(h\cos\phi+v\sin\phi)\right]}, \quad (9)$$

$$\mathbf{M_e} = \frac{1}{\sqrt{2}}\exp(\pm j\phi)\begin{bmatrix}\cos\theta\cos\phi \mp j\sin\phi \\ \cos\theta\sin\phi \pm j\cos\phi \\ \sin\theta\end{bmatrix}, \quad (10)$$

where $f$ is the focal distance, the maximum aperture angle $\alpha = \arcsin(\mathrm{NA}/n_1)$, $\beta$ is the ratio of the pupil radius to the beam waist that we take as unit in the following calculations. $l_1$ and $\mathbf{M_e}$ comes from the contributions of complex amplitude and polarization in the objective space, respectively.

Similarly, the corresponding magnetic field, which is orthogonal to the electric field in homogeneous and isotropic medium, can be described by [57]:

$$\mathbf{H}_i = \frac{-ik_1}{2\pi}\iint_\Omega \mathbf{b}\times\exp[ik_1(\mathbf{s}\cdot\mathbf{r})]d\Omega, \quad (11)$$

$$\mathbf{b} = fl_1\sqrt{\cos\theta}\mathbf{M_m}, \quad (12)$$

$$\mathbf{M_m} = \frac{1}{\sqrt{2}}\exp(\mp j\phi)\begin{bmatrix}-\sin\varphi \pm j\cos\phi\cos\theta \\ \cos\phi \pm j\sin\phi\cos\theta \\ \pm j\sin\theta\end{bmatrix}. \quad (13)$$

Apparently, the distribution of magnetic field in Eqs. (11)–(13) is different from that of the electric field in Eqs. (6)–(10). In terms of the full time-dependent three-dimensional electric and

magnetic fields, the energy current can be defined by the time-averaged Poynting vector [57]:

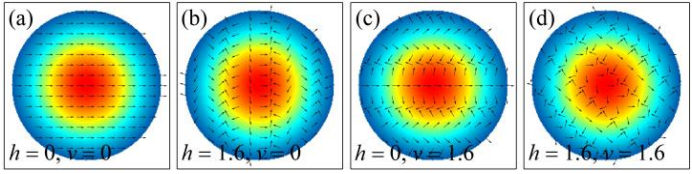

FIG. 3. Polarization distributions of four kinds of SDLLP beams with (a) $(h, v) = (0, 0)$, (b) $(h, v) = (1.6, 0)$, (c) $(h, v) = (0, 1.6)$, and (d) $(h, v) = (1.6, 1.6)$.

$$\mathbf{P} \propto \frac{c}{8\pi} \mathrm{Re}(\mathbf{E}_i \times \mathbf{H}_i^*), \quad (14)$$

where the asterisk gives the operation of the complex conjugation.

For the phase modulated circular polarized $LG_{(0,0)}$ beam mentioned above, which is only carrying SAM per photon. We simulate the electric field distributions in the focal plane with $h = v = 1.6$ when $\delta = \Delta$ as an example. Based on Eqs. (6) and (7), the LH and RH beams will be theoretically focused to locations at $(2.382\lambda, 5\pi/4, 0)$ and $(2.382\lambda, \pi/4, 0)$ in cylindrical coordinate system, respectively. By calculations, we find that the foci are located, respectively, at $(2.382\lambda, 5\pi/4, 0)$ and $(2.382\lambda, \pi/4, 0)$, as depicted in Fig. 2. Obviously, the simulated results agree well with the theoretical predications. The electric field of the longitudinal component has a doughnut-shaped pattern, arriving from the phase singularity of the helical phase in the beam center. To be specific, it is an anticlockwise [Fig. 2(d); for left CP] and clockwise [Fig. 2(i); for right CP] vortex phase distribution, which implies the system transforms the SAM of circular state to OAM state, as reported in Ref. [35]. It is believed that there will exhibit transverse energy flow ring when a light carries OAM, which causes the motion of the trapping particles in optical tweezers, with the trajectory coincides in direction with the Poynting vector, when absorbing a certain portion of the beam energy [59–62]. We simulate the transverse energy flow for the above two beams, as depicted in Figs. 2(e) and 2(j). Obviously, they are ring-shaped patterns with directions in according with the corresponding directions of helical phases, respectively, which further verify the SOC in this system.

When $\delta = \Delta$, we simulate the polarization distributions of four light beams described by Eq. (1) or (2) with $(h, v) = (0, 0)$, (1.6, 0), (0, 1.6), and (1.6, 1.6), shown in Fig. 3. Apparently, the local SoPs are all linearly polarized for all of them, and it is the simply linearly polarized beam with spatially invariant SoP for $(h, v) = (0, 0)$. However, for other values of $(h, v)$, the local SoP is spatially variant along a series of parallel lines, in contrast to the well-known radially or azimuthally polarized beam, for which the local SoP is spatially variant along a series of concentric rings. The angel for the parallel lines with respect to $x$ axis can be given as $\phi = \arccos\left(\dfrac{h}{\sqrt{h^2+v^2}}\right)$. For the direction perpendicular to these lines, the local SoP is spatially invariant. They are new members of VOF, we here call them as LVLLP also consider the $LG_{(l, p)}$ laser mode and chose $p = 0$, for the input light described by Eqs. (1) or (2), the corresponding electric field strength factor **a** in Eq. (3) and magnetic field strength factor **b** in Eq. (11) in the image space can be derived, respectively, as

$$\mathbf{a} = fl_0\sqrt{\cos\theta}\mathbf{M}_e, \quad \mathbf{b} = fl_0\sqrt{\cos\theta}\mathbf{M}_m \quad (15)$$

$$l_0 = \left(\sqrt{2}\beta\frac{\sin\theta}{\sin\alpha}\right)^{|l|}\exp\left[-\left(\beta\frac{\sin\theta}{\sin\alpha}\right)^2\right]\exp(jl\phi), \quad (16)$$

$$\mathbf{M}_e = \begin{bmatrix} \sin(\phi-\delta)\sin\phi + \cos(\delta-\phi)\cos\theta\cos\phi \\ -\sin(\phi-\delta)\cos\phi + \cos(\delta-\phi)\cos\theta\sin\phi \\ \cos(\delta-\phi)\sin\theta \end{bmatrix}, \quad (17)$$

$$\mathbf{M}_m = \begin{bmatrix} -\sin\phi\cos(\delta-\phi) + \cos\phi\cos\theta\sin(\phi-\delta) \\ \cos\phi\cos(\delta-\phi) + \sin\phi\cos\theta\sin(\phi-\delta) \\ \sin\theta\sin(\phi-\delta) \end{bmatrix}. \quad (18)$$

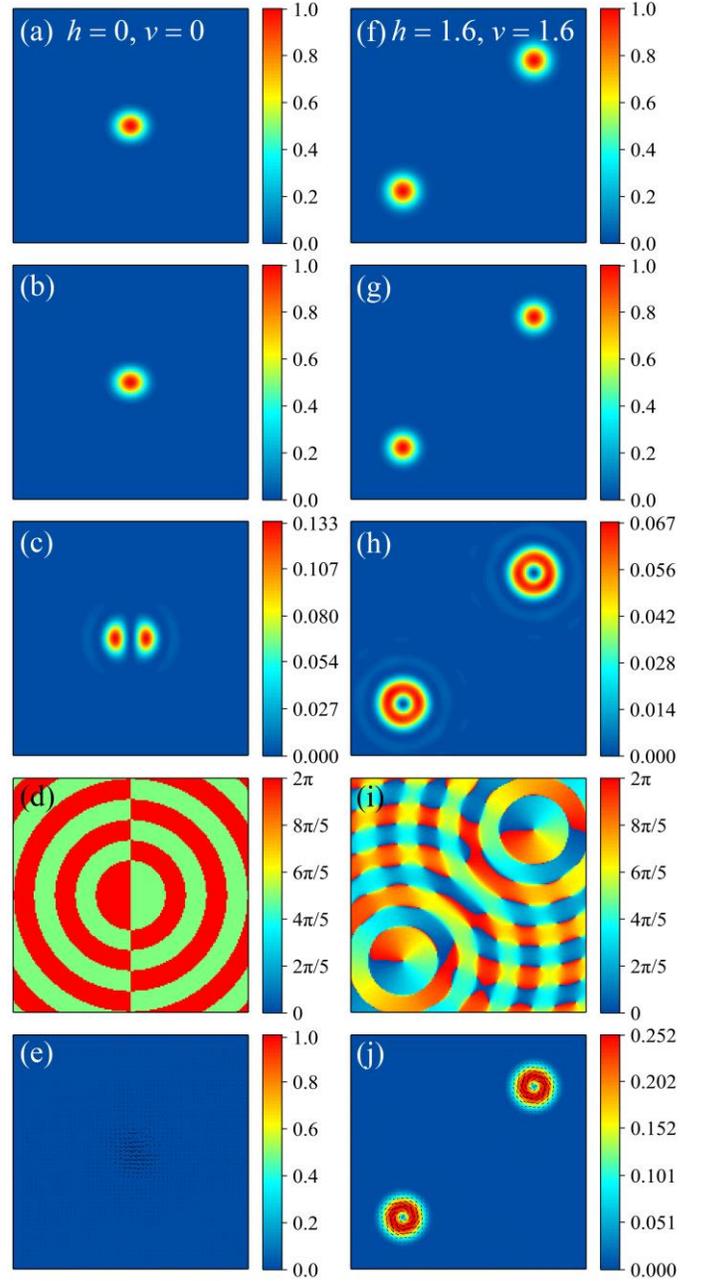

FIG. 4. Simulated electric field and energy flow of tightly focused LVLLP $LG_{(0,0)}$ beams in the focal plane with $(h, v) = (0, 0)$ (left column) and $(1.6, 1.6)$ (right column). (a),(f) Total intensity distribution. (b),(g) Intensity distribution of the transverse component. (c),(h) Intensity distribution of the longitudinal component. (d),(i) Phase distribution of the electric field of the longitudinal component. (e),(j) Energy flow of the transverse component, the direction of which is shown by the black arrows. The size for each image is $6\lambda \times 6\lambda$.

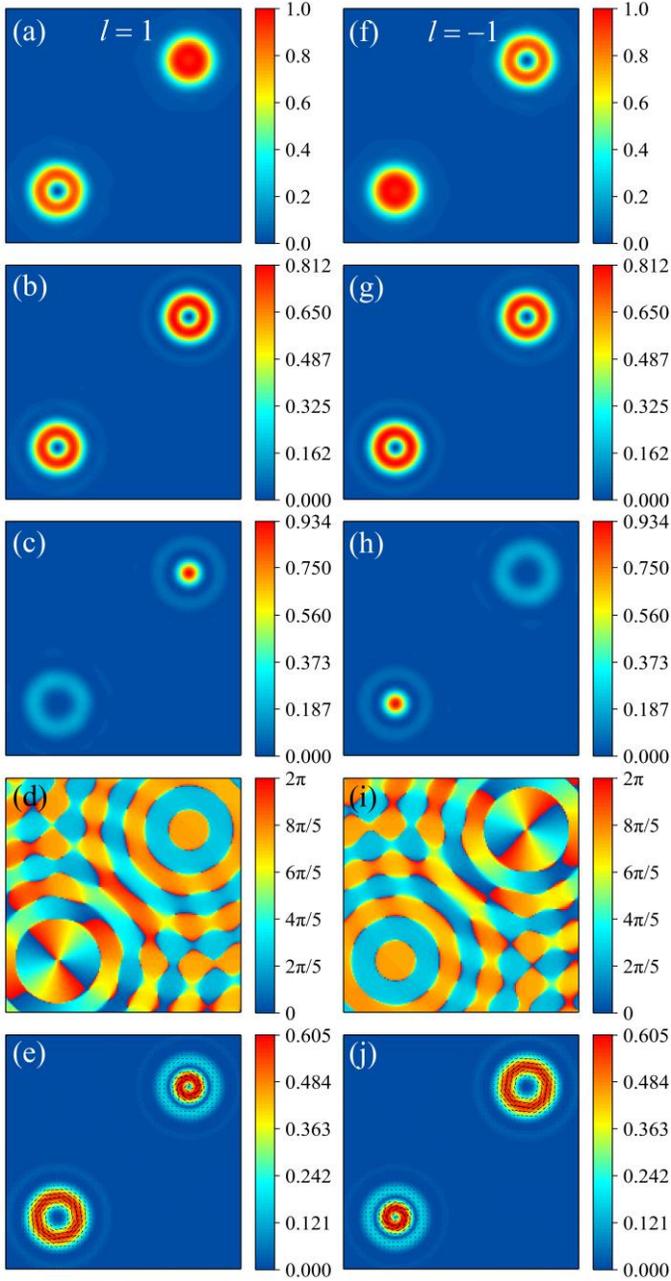

For the LVLLP $LG_{(0, 0)}$ beam, which is not carrying any SAM and OAM per photon. For the sake of comparison, we simulate the electric field in the focal plane of two LVLLP $LG_{(0, 0)}$ beams with $(h, v) = (0, 0)$ and $(1.6, 1.6)$, respectively, depicted in Fig. 4. For $(h, v) = (0, 0)$, the foci for the two components are completely overlapping, based on Eq. (5). As a result, we here can only obtain a single focus for the total field [Fig. 4(a)]. They are twin hot spots with semicircular-shaped patters for the longitudinal component [Fig. 4(c)], arriving from the phase singular line along $y$ axis [Fig. 4(d)]. In this case, there is no helical phase related to OAM, and only exhibits a $\pi$ shift in the two side of this line [Fig. 4(d)]. Further, on transverse energy flow can be found, as shown in Fig. 4(e). To the contrary, for $(h, v) = (1.6, 1.6)$, the foci for the two components are completely separated as shown in Fig. 2(c). Hence, we here gain similar double foci for the total, transverse, and longitudinal fields [Figs. 4(f)–4(h)]. The top right focus is mainly contributed by the RH circular component, while bottom left focus is mainly contributed by the LH circular component. Most importantly, there exhibit double helical phases with opposite direction for the longitudinal component of the electric field [Fig. 4(i)], which imply the system transformed linear state to conjugate OAM states. Thus, double ring-shaped transverse energy flows with opposite direction play the dominate role, shown in Fig. 2(j).

To further prove the partial OAM conversion of light without spin state, we consider LVLLP-$LG_{(1, 0)}$ and LVLLP-$LG_{(-1, 0)}$ beams with $(h, v) = (1.6, 1.6)$, which carry none SAM and, respectively, $+\hbar$ and $-\hbar$ OAM per photon. For both cases, the double foci are much different from each other for their total electric fields [Figs. 5(a) and 5(f)], arriving from the peculiar distributions of the longitudinal component [Figs. 5(c) and 5(h)]. To be specific, there is no helical phase related to OAM for the top right [bottom left] focus in the longitudinal component,

FIG. 5. Simulated electric field and energy flow of tightly focused LVLLP $LG_{(1, 0)}$ (left column) and $LG_{(-1, 0)}$ (right column) beams in the focal plane with $(h, v) = (1.6, 1.6)$. (a),(f) Total intensity distribution. (b),(g) Intensity distribution of the transverse component. (c),(h) Intensity distribution of the longitudinal component. (d),(i) Phase distribution of the electric field of the longitudinal component. (e),(j) Energy flow of the transverse component, the direction of which is shown by the black arrows. The size for each image is $6\lambda \times 6\lambda$.

beams, and linearly polarized beam can be seen as one extreme case of LVLLP beams.

Theoretically, when the two components of a light can be focused to different locations, we should obtain twin foci for this light. One focus is contributed by one component, while the other is contributed by the other component. To verify it, we depicted in Fig. 5(d) [Fig. 5(i)], which is caused by the compensation of the opposite OAM arriving from OAM conversion of optical field from linear state into conjugate OAM states. In contrary, double helical phase appears for the bottom left [top right] focus shown in Fig. 5(d) [Fig. 5(i)] when the directions of the input and converted OAMs are the same. These phase changes are manifested in the intensity profiles; they are bright spots for the top right [bottom left] focus of the longitudinal and total fields depicted, respectively, in Figs. 5(c) and 5(a) [Figs. 5(h) and 5(f)], while for the bottom left [top right] focus, however, they are donut-shaped patters for longitudinal and total fields shown, respectively. All the foci for the transverse component for both cases exhibit doughnut-shaped patters are due to the modulation of input OAM.

Chirality describes the geometric property of an object and makes it to be distinguished from its mirror image. As a common phenomenon in natural systems, chirality is ubiquitous, especially in many biochemical molecules, such as sugars, proteins, and amino acids. The chirality of a molecule is strongly linked to biological function, which means that the molecules with opposite chirality can exhibit completely different bioactivities to organisms. Once a chiral protein lost its original chirality, it may be toxic to cells and potentially cause some diseases [63]. Therefore, the identification and separation of substances by chirality have always been a research hotspot in biochemical and pharmaceutical industries [64,65]. According to the most recent findings, light can be one of the efficient and potential candidates to separate chiral objects, since the particles with opposite chiralities tend to interact differently with the same light field [66–69]. The spherical chiral particle with a radius much smaller than the wavelength can be regard as a pair of interacting elelctric and magnetic dipoles under the excitation of an external time-harmonic electromagnetic field. In this case, the time-averaged optical force exerted on the chiral dipoles can be written as [70]:

$$\langle \boldsymbol{F} \rangle = \frac{1}{2}\mathrm{Re}\left[\mathbf{p}\cdot(\nabla)\mathbf{E}+\mathbf{m}\cdot(\nabla)\mathbf{H}+\frac{ck_1^4}{6n_1\pi}\left(\mathbf{p}\times\mathbf{m}^*\right)\right], \quad (20)$$

where $k_1 = n_1\omega/c$ is the wave number with $n_1 = (\varepsilon_1\mu_1)^{1/2}$ being the refractive index of background medium, $\omega$ being the angular frequency of external harmonic field, and $c$ being the light speed in vaccum. $\mathbf{p}$ and $\mathbf{m}$ are the induced electric and magnetic moments, which can be described by the following relations [71]:

$$\mathbf{p} = \alpha_e\mathbf{E} + i\alpha_c\mathbf{H}, \quad (21)$$

$$\mathbf{m} = \alpha_m\mathbf{H} - i\alpha_c\mathbf{E}, \quad (22)$$

where $\alpha_e$, $\alpha_m$, and $\alpha_c$ denote, respectively, the electric polarizability, the magnetic polarizability, and the chiral polarizability of the particle, which are the complex functions of the relative permittivity $\varepsilon_2$, the relative permeability $\mu_2$, and the chirality parameter $\kappa$. Using the notations of Bohren and Huffman [72], one can derive the three polarizabilities of the particle from its Mie scattering cofficients as:

$$\alpha_e = \frac{i6\pi\varepsilon_1\varepsilon_0}{k_1^3}a_1, \quad \alpha_m = \frac{i6\pi\mu_1\mu_0}{k_1^3}b_1, \quad \alpha_c = \frac{6\pi n_1}{ck_1^3}c_1, \quad (23)$$

here $\varepsilon_0$ and $\mu_0$ are the permittivity and permeability of vacuum, separately. $a_1$, $b_1$, and $c_1$ are the corresponding Mie scattering cofficients within the small particle limit.

Based on Eqs. (20)–(23), the force distribution subjected by a spherical chiral particle in the above tightly focused vector beams can be accurately calculated. In the following simulations, a pair of chiral particles with the radius $a = 40$ nm, the relative permittivity $\varepsilon_2 = 2.5$, the relative permeability $\mu_2 = 1$, and the opposite chiralities $\kappa = 1$ and $-1$ are considered. Besides, we assume that the input power P = 100 mW, the incident wavelength $\lambda_0 = 1.064$ μm, the numerical-aperture NA = 1.2, the refractive index $n_1 = 1.33$, and the relative permeability $\mu_1 = 1$. Fig. 6 gives the transverse force distributions in the focal plane exerted on the pair of chiral particles under the excitation of the tightly focused LVLLP $LG_{(0, 0)}$ beam with $(h, v) = (0.6, 0.6)$, where the green arrows and background show the direction and magnitude of the force, respectively. As depicted in Fig. 6(a), for a right-handed particle with chirality parameter $\kappa = 1$, the transverse force at any position of bottom left focus is pointing to the center of focal spot, while the transverse force at any point of top right focus is pointing to the outer of focal spot. However, the scenario is dramatically different for a left-handed particle with chirality parameter $\kappa = -1$, as shown in Fig. 6(b). It can be seen that the transverse force will change its sign as the chirality of particle is reversed. Consequently, for a right-handed [left-handed] particle, the transverse force pushes it toward the center of bottom left [top right] focus enabling a stable transverse trapping, but pushes it away from the center of top right [bottom left] focus plane manifesting a non-trapping phenomenon.

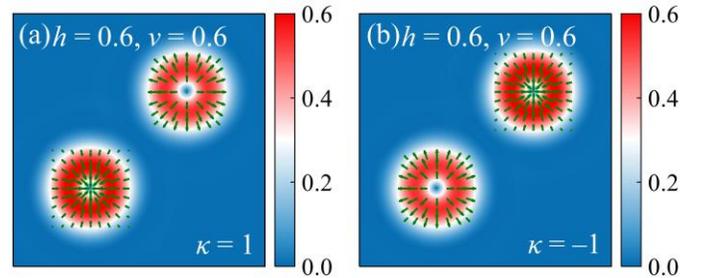

FIG. 6. Calculated transverse force distributions subjected by a pair of chiral nanoparticles with the opposite chiralities (a) $\kappa = 1$ and (b) $\kappa = -1$ in the tightly focused field of a LVLLP $LG_{(0, 0)}$ beam with $(h, v) = (0.6, 0.6)$. The direction of the force is indicated by the green arrows. The size for each image is $2.6\lambda \times 2.6\lambda$.

To summarize, we have demonstrated that OAM conversion of light without any spin state is possible in a homogenous and isotropic medium, provided the LVLLP beam is strongly focused. In such a strong focusing system, LVTLPs can generate identical twin foci with tunable distances between them controlled by the input SoPs for all the transverse, longitudinal, and total fields in the focal region. Most importantly, such system achieves partial OAM conversion from linear state to conjugate OAM states, giving rise to helical phases with opposite directions for each focus in the longitudinal component of the electric field. Due to such OAM conversion, there exhibits twin transverse energy flow rings with opposite directions in the focal plane. Furthermore, such customized light field can be used to identify and separate chiral objects in optical tweezers.

This work was partially supported by National Natural Science Foundation of China (NSFC) (11604182, 11704226); Natural Science Foundation of Shandong Province (ZR2016AB05, ZR2017MA051).

* zsman@sdut.edu.cn